\documentclass[doublecol]{epl2} 
\usepackage[T1]{fontenc}

\title{Making two dysprosium atoms rotate - Einstein-de Haas effect revisited}
\shorttitle{Einstein-de Haas effect for two dysprosium atoms} %Insert here a short version of the title if it exceeds 70 characters

\author{Wojciech G\'orecki\inst{1} \and Kazimierz Rz\k{a}\.zewski\inst{1}}
\shortauthor{Wojciech G\'orecki \etal}

\institute{                    
  \inst{1} Center for Theoretical Physics, Polish Academy of Sciences - Al. Lotnik\'ow 32/46, 02-668 Warsaw, Poland\\
}
\renewcommand{\b}{\textbf}
\usepackage{amssymb}
\usepackage{amsmath}

\abstract{We present a numerical study of the behaviour of two magnetic dipolar atoms trapped in a harmonic potential and exhibiting the standard Einstein-de Haas effect while subject to a time dependent homogeneous magnetic field. Using a simplified description of the short range interaction and the full expression for the dipole-dipole forces we show, that under experimentally realisable conditions two dysprosium atoms may be pumped to a high ($l>20$) value of the relative orbital angular momentum.}

\begin{document}

\maketitle

\section{Introduction}
The Einstein-de Haas effect (EdH) \cite{edh} is a classic physical phenomenon discovered more than 100 years ago. The magnetism of a ferromagnet is associated with the electronic spins. Due to the angular momentum conservation, freely suspended piece of ferromagnet starts to rotate when its magnetization is changed because of the changing external magnetic
field. The effect has been experimentally demonstrated by the authors placing a ferromagnetic cylinder inside a coil and driving a burst of electric current through this coil.\par
As the angular momentum conservation is universal, the effect occurs in micro-scale as well. Recently the first experimental observations of the EdH effect was reported for a single molecule \cite{eksperyment}. Simultaneously, since a successful condensation of chromium \cite{chrom1,chrom2}, due to its large magnetic dipole moment, the interest in the EdH was also extended to gaseous Bose-Einstein condensates \cite{EdH1,EdH2}.\par
As in the original idea of Einstein and de Haas, the magnetic dipole interaction couples the atomic spins to the orbital angular momentum. Such a coupling is the essential ingredient of the effect. With the condensation of erbium \cite{erbium} and dysprosium \cite{dysp1,dysp2} even  stronger dipolar forces became available. For chromium \cite{chredh} and rubidium \cite{rubid1,rubid2,rubid3} condensates the EdH effect has been investigated. With an eye on experiments in deep optical lattices \cite{opty1,opty2} several theoretical papers investigated the simplest multi atom system exhibiting EdH effect - a two atom system \cite{2atomy1,2atomy2}.
\par
In this Letter we analyze the EdH effect for two dysprosium atoms placed in a spherical harmonic trap. The dysprosium fermionic isotope has its spin equal $(21/2)\hbar$. Hence a significant computational complexity of the problem at hand. Here we focus on the time evolution of the ground state while the value of the homogeneous magnetic field is gradually changed. In contrast to \cite{2atomy1,2atomy2} we do not use the perturbation theory; the calculation method is very similar to the one from \cite{my}. It turns out that the large effect could be reached for available magnetic field strength and realistic time of changing this field. A lot of orbital angular momentum may be generated in the system.

\section{Theoretical model}
Let us consider a system of two identical dipolar atoms in an isotropic harmonic trap with external homogeneous magnetic field $\b{B}$. The Hamiltonian is given as:
\begin{equation}
H
=H_{OSC}(\b{r}_1,\b{r}_2)
+H_{INT}(\b{r}_1-\b{r}_2,\b{F}_1,\b{F}_2)
+H_{B}(\b{F}_1+\b{F}_2)
\end{equation}
where $\b{r}_1$, $\b{r}_2$ are the position vectors of the atoms and $\b{F}_1$, $\b{F}_2$ are their spins (total internal angular momenta). Here $H_{OSC}=-\frac{\hbar}{2m}\nabla^2_1+\frac{1}{2}m\omega^2r^2_1-\frac{\hbar}{2m}\nabla^2_2+\frac{1}{2}m\omega^2r^2_2$, $H_{INT}$ is an interaction energy (which contains both short range ($H_{SR}$) and long range dipole-dipole ($H_{DD}$) parts) and $H_{B}$ is the energy of the magnetic dipoles in an external magnetic field.\par
In $H_{OSC}$ one can separate the part governing the mass center motion $H_{CM}$ and the part describing a relative motion of atoms $H_{REL}$ by introducing:
$\b{r}=\frac{1}{\sqrt{2}}(\b{r}_1-\b{r}_2)$, $\b{R}=\frac{1}{\sqrt{2}}(\b{r}_1+\b{r}_2)$:
\begin{multline}
H_{OSC}=H_{CM}+H_{REL}=\\
(-\frac{\hbar}{2m}\nabla^2_R+\frac{1}{2}m\omega^2R^2)+
(-\frac{\hbar}{2m}\nabla^2_r+\frac{1}{2}m\omega^2r^2)
\end{multline}
One can see that the center of mass motion separates and is independent of the relative motion.
We assumed $H_{SR}$ to be spherically symmetric $H_{SR}(\b{r})=H_{SR}(r)$ (more about that in the \b{Results}). $H_{DD}$ is given as:
\begin{equation}
H_{DD}=\frac{\mu_0(\mu_{B}g_j)^2}{4\pi|\sqrt{2}\b{r}|^3}[\b{F}_1\cdot\b{F}_2-3(\b{F}_1\cdot\b{n})(\b{F}_2\cdot\b{n})]
\end{equation}
where $\b{n}=\frac{\b{r}}{|\b{r}|}$, $\mu_{0}$ is the vacuum magnetic permeability, $\mu_B$ is the Bohr magneton, $g_j$ is Lande g-factor and $\b{F}_1$, $\b{F}_2$ are %dimensionless
spin operators. Spin-statistic theorem says, that in a system of two identical particles, $l$ and $f$ have the same parity (both are even or both are odd). In addition, obviously, the highest $f$ is even (odd) for bosons (fermions). It can be easily proved that if a state has the well defined orbital angular momentum $\b{L}$ and spin $\b{F}$, acting $H_{DD}$ does not change the parity of $\b{L}$ and $\b{F}$.  What is more, $[H_{DD},\b{J}]=0$ (where $\b{J}=\b{L}+\b{F}$). At last (we assume that $\b{B}=(0,0,B)$):
\begin{equation}
H_{B}=-g_{j}\mu_B\b{B}\cdot \b{F}=-g_{j}\mu_BBF_z
\end{equation}
where $F_z=F_{1z}+F_{2z}$ (external field is directed along the z-axis). $[H_{B},J_z]=0$, but generally $[H_{B},\b{J}]\neq0$. Thus, eigenstates of the system have the well defined $m_j$ and parity of $\b{L}$ and $\b{F}$ (which will be denoted by p=o/e in the superscript). The eigenstate can be written as:
\begin{equation}
\Psi^{m_jp}_n=\sum\limits_{f,m_f,l}\phi^{m_jfm_fl}_n(r)|f,m_f,l,m_l=m_j-m_f\rangle
\end{equation}
Where $\phi^{m_jfm_fl}_n(r)$ are radial functions.
$f$ is limited because $f\leqslant |f_1+f_2|$, but $l$ is not (the only restriction is $l\geqslant|m_j-m_f|$), so the sum has infinitely many terms. Let us remind the reader, that for a pure harmonic oscillator Hamiltonian, the eigenvalues are given as $E_{nl}=\hbar\omega(2n+l+\frac{3}{2})$, so the energy grows linearly with $l$. As we expect, that eigenstates of our system are combinations of functions, which are quite similar to the eigenstates of a free oscillator and we are interested in a few lowest states, it is reasonable to cut the sum over $l$ at a suitable value $l_{max}$. This value is determined by numerical tests.
\begin{equation}
\Psi^{m_jp}_n=\sum\limits_{f,m_f,l}^{f_1+f_2,l_{max}}\phi^{m_jfm_fl}_n(r)|f,m_f,l,m_l=m_j-m_f\rangle
\end{equation}
Thus we arrive at a finite system of equations for radial functions $\phi^{m_jpfm_fl}_n(r)$, which we can write down using harmonic oscillator units $q=\sqrt{\frac{m\omega}{\hbar}}r$ and introducing  dimensionless $g_{dd}=\frac{\mu_0(\mu_{B}g_j)^2}{8\sqrt{2}\pi}\sqrt{\frac{m^3\omega}{\hbar^5}}$, $b=\frac{g_j\mu_{B}}{\hbar\omega}B$, $\chi^{m_jfm_fl}_n(q)=q\phi^{m_jfm_fl}_n(q)$:
\begin{multline}
-\frac{1}{2}\frac{d^2}{dq^2}\chi^{m_jfm_fl}_n(q)+\frac{1}{2}q^2\chi^{m_jfm_fl}_n(q)+\\+\frac{l(l+1)}{2q^2}\chi^{m_jfm_fl}_n(q)+H_{SR}(q)+
\\+\frac{g_{dd}}{q^3}\sum\limits_{f',m_f',l'}\alpha_{fm_fl,f'm_f'l'}\chi^{m_j'f'm_f'l'}_n(q)+\\-bm_f\chi^{m_jfm_fl}_n(q)=E^{m_jp}_n\chi^{m_jfm_fl}_n(q)
\end{multline}
where $\alpha_{fm_fl,f'm_f'l'}$ coefficients are the matrix elements:
\begin{multline}
\alpha_{fm_fl,f'm_f'l'}=\\=\langle fm_flf'm_{f'}l'|[\b{F}_1\cdot\b{F}_2-3(\b{F}_1\cdot\b{n})(\b{F}_2\cdot\b{n})]|fm_flf'm_{f'}l'\rangle
\end{multline}

\begin{figure*}[ht]
\begin{center}
\includegraphics[width=0.8\textwidth]{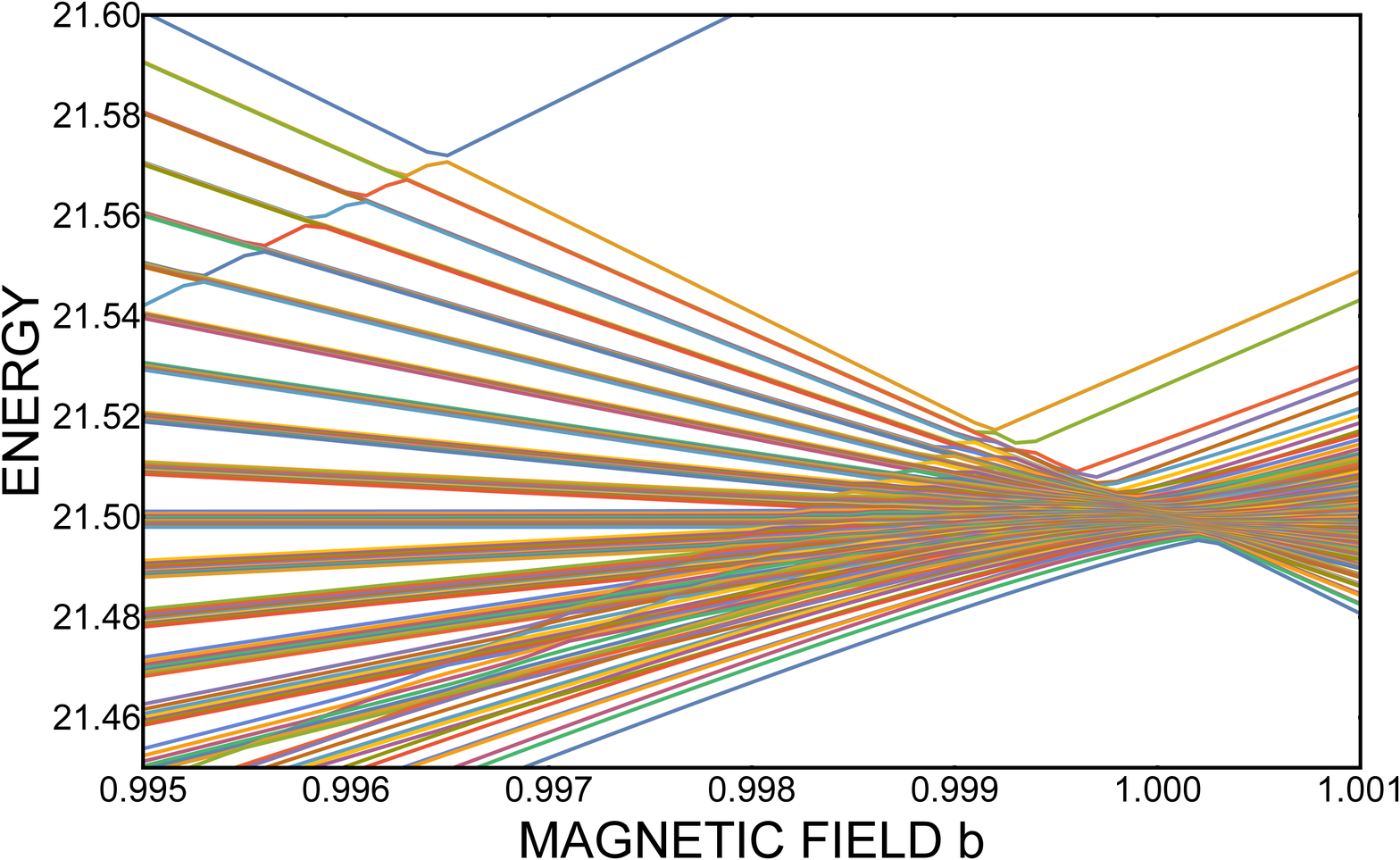}
\caption{Energies of eigenstates $vs.$ magnetic field. Anticrossings of $121$ eigenstates occur for $B\approx\frac{\hbar\omega}{g_j\mu_b}$ ($b\approx 1$). Strong asymmetry can be explained by the fact that the short range interaction increases eigenenergies of states with low $\langle L^2 \rangle$ value and nearly does not change the energy of states with high $\langle L^2 \rangle$.}
\label{fig:ev}
\end{center}
\end{figure*}

\begin{figure*}[ht]
\begin{center}
\includegraphics[width=0.8\textwidth]{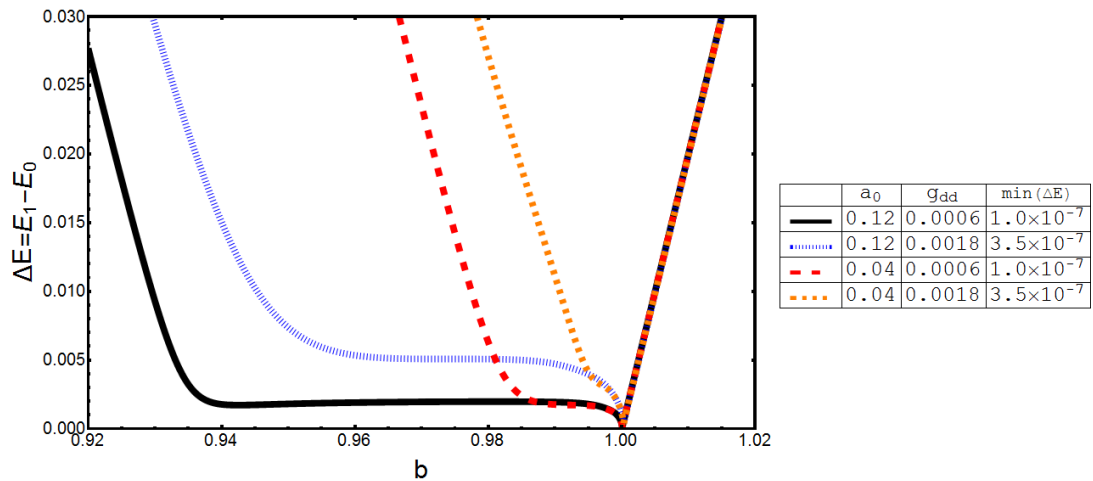}
\caption{Difference between energies of the first excited states and the ground state for four cases. One can notice that the $b$ value, for which curve start to be flat, depends strongly on $a_0$, and the size of gap is determined by $g_{dd}$ value.}
\label{fig:gap}
\end{center}
\end{figure*}

\begin{figure*}[ht]
\begin{center}
\includegraphics[width=0.9\textwidth]{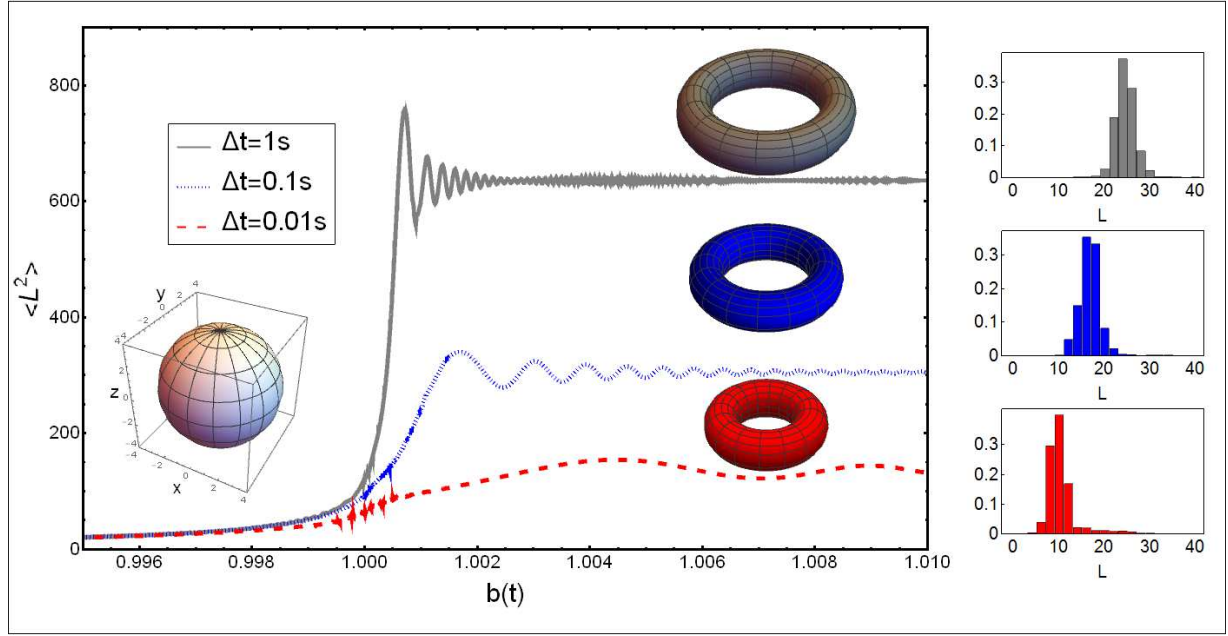}
\caption{Time evolution of the initial ground state for three different times of transition through the anticrossing region. Time $\Delta t$ is the time of changing the magnetic field value $b$ from $-0.99$ to $1.01$. On the main chart the distribution of atoms' relative position is presented -- at the beginning (left, the same for all cases) and at the end of the evolution. On the right there are histograms of various orbital momentum contributions to the final state}
\label{fig:wh}
\end{center}
\end{figure*}

\begin{figure*}[ht]
\begin{center}
\includegraphics[width=1\textwidth]{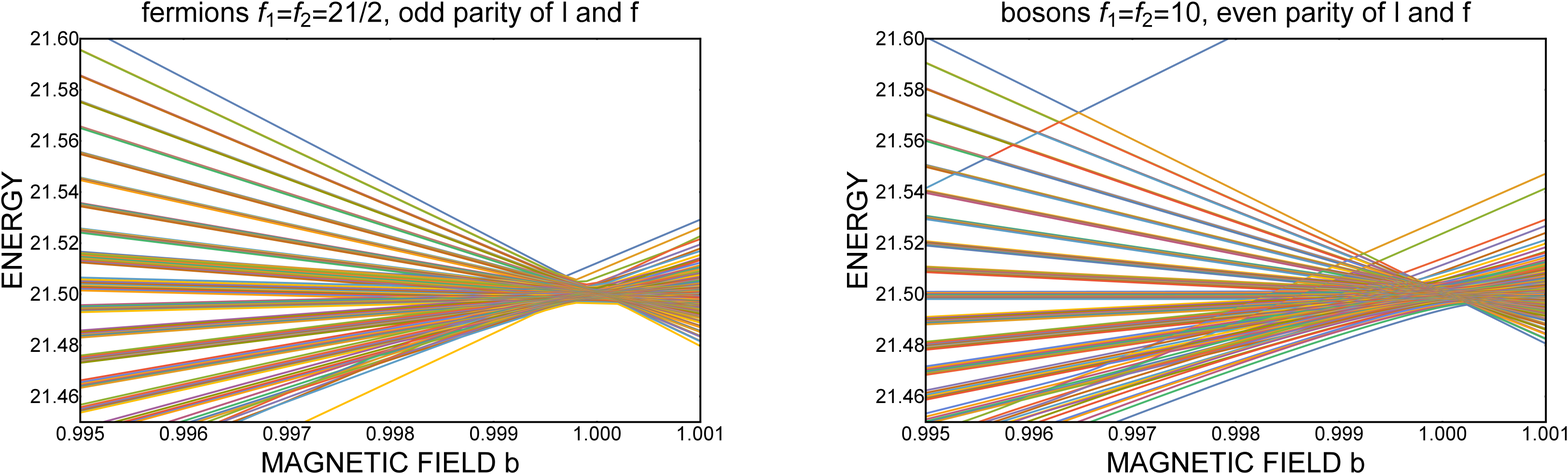}
\caption{
Energies of eigenstates $vs.$ magnetic field for two other cases. The plot for fermions with odd parity of $l$ and $f$ is quite different than the previous one (fig.\ref{fig:ev}) - here there is no eigenstate with energy strongly increased by the short range interaction (no state $l=0$). On the other hand, the plot for bosons is very similar to the dysprosium case.
}
\label{fig:evc}
\end{center}
\end{figure*}

\section{Results} We consider the case of fermionic dysprosium atoms with total internal angular momentum $f_1=f_2=\frac{21}{2}$ in a harmonic trap of frequency $2\pi\cdot 30kHz$, for which $g_{dd}=0.0018$. For calculations we assume that $H_{SR}(q)$ potential takes a simple form:
\begin{equation}
    H_{SR}(q):= \begin{cases}
               +\infty               & for\quad q<a_0=89r_0\\
               0               & for\quad q>a_0=89r_0\\
           \end{cases}
\end{equation}
where $r_0$ is the Bohr radius and $a_0$ is the s-wave scattering length \cite{kontakt}; its value in dimensionless harmonic oscillator units is $a_0=0.12$. We do it for a few reasons: firstly, although the short range potential is rather complicated, in cold atoms experiments only the s-wave scattering length matters so realistic radius-dependence is not essential.
It is known that the s-wave scattering lengths for dysprosium \cite{kontakt} are different in different scattering channels. With its precise values not known, we are simplifying our task by assuming only one, universal scattering length $a_0$. 
What is more, this approach lets  us neglect the electric interaction (which are of quadrupole-quadrupole type or weaker). At last, we are interested in cases when $g_j\mu_BB\approx\hbar\omega\rightarrow B\approx\frac{\hbar\omega}{g_j\mu_b}\approx0.1G$. This value of $B$ is far away from magnetic Feshbach resonances \cite{rez}, so it is justified to assume, that value of $a_0$ does not depend on $B$.\par
Generally, the interaction energy $H_{DD}$ and $H_{SR}$ is much smaller than the others, so for most $b$ values the eigenenergies are very close to energies of sates $|n,f,m_f,l,m_l\rangle$, where $n$ is the harmonic oscillator number. For states with $l>0$ any changes generated by $H_{SR}$ are practically negligible. For s-states change of the shape of eigenfunction is significant, but the value of energy shift is much smaller than $\hbar\omega$. Therefore, the energy in a very crude approximation can be written as:
\begin{equation}
E^{fm_flm_l}_{n}\approx(\frac{3}{2}+2n+l)-bf_j
\end{equation}
(Note, that we used this approximation only to find the value of $b$, for which interesting effects occur. Of course, in our calculations we include all interactions.)
One can note, that for $b=1$, every state $|0,f,m_f,m_f+20,-(m_f+20)\rangle$ (in absence of dipole-dipole and short range interactions) would have the same energy $\frac{43}{2}$. It means that here anticrossings \cite{landau,zener} (generated by $H_{DD}$ interaction) must occur. 
\par
In fig.\ref{fig:ev} we plotted the eigenenergies of the full Hamiltonian as a function of dimensionless magnetic field $b$. One can note that the chart is definitely asymmetrical for reflection with respect to the vertical $b=1$ line. The reason for this is the property of short range interaction. This interaction increases the energy of states with low $l$ values and almost does not change the energy of states with high $l$ -- in the second case the atoms do not reach the hard sphere because of the strong centrifugal barrier. This is the reason for observed asymmetry.\par
In fig.\ref{fig:gap} we plotted the value of a difference between eigenenergies of the first excited state and the ground state (for broader range of $b$ than in the previous plot). Besides our case (blue, tiny dashed curve), we have considered here three other examples (note that the red dashed curve ($a_0=0.04,g_{dd}=0.0006$) responds to dysprosium atoms in trap with frequency  $2\pi\cdot 3kHz$; two others are purely hypothetical). We can observe two effects: in every single case, for some range of $b$, $\Delta E$ is flat. At the beginning of this range anticrossing between state with $l=0$ and the lowest state with $l=2$ occurs. The energy gap here is really broad, so it is not significant for further considerations. One can notice, that position of this point depends strongly on $a_0$ value - indeed, for higher $a_0$ the energy of the ground state is more up-shifted, so it will meet the first excited stated for lower $b$ value. What will be really important is the very narrow energy gap for $b$ value close to $1.00$. The size of the gap depends only on $g_{dd}$ value.
Now let us consider the evolution which leads to EdH effect.  It is fully determined by the Schr\"ondinger equations written in the time dependent bases \cite{adiabatic}:
\begin{align}
\begin{split}
\Psi(t)&=\sum_n c_n(t)\psi_n(t)e^{i\theta_n(t)}\\
\theta_n(t)&=-\frac{1}{\hbar}\int\limits_0^t E_n(t')dt'\\
\dot{c}_m(t)&=-c_m\langle \psi_m|\dot{\psi}_m\rangle-\sum_{n\neq m}c_n\frac{\langle\psi_m|\dot{H}|\psi_n\rangle}{E_n-E_m}e^{i(\theta_n-\theta_m)}
\end{split}
\end{align}
We start with $b=-0.2$. The ground state of the whole system is $|0,20,-20,0,0\rangle$, so the parity of $l$ and $f$ (even) and the value $m_j=-20$ are fixed for further evolution. 
If the fully adiabatic transition was possible the maximal value of $l=40$ could be reached. The time scale for the perfect adiabatic transition is of the order of a reciprocal value of the gap at the lowest aticrossings and it is longer than 15s. %It is unrealistically long.
Therefore we looked at the evolution for the realistic time scales.
In fig.\ref{fig:wh} we present the time evolution for three cases. $\Delta t$ is the time of changing magnetic field $b$ from value $0.99$ to $1.01$ - time of changing magnetic field from $-0.2$ to $0.99$ can be much shorter, because there is no narrow anticrossing for this $b$ values (reciprocal of difference between energies of the two lowest states is bigger than $0.001s$), so the state remains nearly unchanged. In the case of $\Delta t=1.0s$, $\langle L^2\rangle$ as large as $600$ is achieved. As is shown at the right panel in fig.\ref{fig:wh}, the dominant contributions to the final state in this case are for $20\leq l \leq 30$. Hence, the relative motion in this case is confined to a relatively thin ring.\par

In fig.\ref{fig:evc} we plotted the eigenenergies for two other cases: for odd parity of $l$ and $f$ (instead of even) and for bosons (instead of fermions). One can notice that changing parity of $l$ and $f$ makes significant modification of the $E(b)$ relation - in case of odd parity there is no state with energy highly increased by short range interaction. The plot is much closer to be symmetrical for reflection with respect to the vertical $b=1$ line. On the other hand, when we consider two bosons instead of two fermions (with similar value of spin), the plot looks almost the same as in the previous case (fig.\ref{fig:ev})

\section{Conclusions} We have presented a simple calculation of the Einstein-de Haas effect for a system of two dysprosium atoms trapped in a spherically symmetric harmonic potential and subject to a time dependent magnetic field. The dipole-dipole interaction couples the spin variables to the orbital angular momentum of the relative motion. A large spin of dysprosium atoms allows to pump a significant amount of the orbital angular momentum into the system. The narrow anticrossings limit the highest achievable $l$ if we limit ourselves to realistic time scales of the magnetic field switching.

\acknowledgments
The authors are pleased to acknowledge fruitful conversations with R. O\l{}dziejewski, M. Gajda and M. Brewczyk.  This work was supported by the (Polish) National Science Center Grant No. DEC-2012/04/A/ST2/00090.


\begin{thebibliography}{0}

%\bibitem{b.a}
 % \Name{Author F., Author S. \and Author T.}
 % \REVIEW{Some Rev. A}{69}{1969}{9691}.

%\bibitem{b.b}
 % \Name{Author F. \and Author S.}
  %\Book{Some Book of Interest}
  %\Editor{A. Editor}
  %\Vol{9}
  %\Publ{Publishing house, City}
  %\Year{1939}
  %\Page{666}.

%\bibitem{b.c}
  %\Editor{Editor A.}
  %\Book{Some Book of Interest}
  %\Vol{9}
  %\Publ{Publishing house, City}
  %\Year{1939}
  %\Section{A}.
  
\bibitem{edh} 
\Name{Einstein A. \and de Haas W. J.}
\REVIEW{Verh. Dtsch. Phys. Ges.}{17}{1915}{152}.

\bibitem{eksperyment} 
\Name{Ganzhorn M., Klyatskaya S.,Ruben M. \and Wernsdorfer W.}
\REVIEW{Nature Communication}{7}{2016}{11443}.

\bibitem{chrom1} 
\Name{Beaufils Q., Chicireanu R., Zanon T., Laburthe-Tolra B., Mar\'echal E., Vernac L., Keller J.-C. \and Gorceix O.}
\REVIEW{Phys. Rev. A}{77}{2008}{061601(R)}.

\bibitem{chrom2} 
\Name{Griesmaier A., Werner J., Hensler S., Stuhler J. \and Pfau T.}
\REVIEW{Phys. Rev. Lett.}{94}{2005}{160401}.


\bibitem{EdH1} 
\Name{Pasquiou B., Mar\'echal E., Bismut G., Pedri P., Vernac L., Gorceix O. \and Laburthe-Tolra B.}
\REVIEW{Phys. Rev. Lett.}{106}{2011}{255303}.

\bibitem{EdH2} 
\Name{Lahaye T., Menotti C., Santos L., Lewenstein M. \and Pfau T.}
\REVIEW{Rep. Prog. Phys.}{72}{2009}{126401}.


\bibitem{erbium} 
\Name{Aikawa K., Frisch A., Mark M., Baier S., Rietzler A., Grimm R. \and Ferlaino F.}
\REVIEW{Phys. Rev. Lett.}{108}{2012}{210401}.

\bibitem{dysp1} 
\Name{Lu M., Burdick N. Q., Youn S. H. \and Lev B. L.}
\REVIEW{Phys. Rev. Lett.}{107}{2011}{190401}.

\bibitem{dysp2} 
\Name{W{\"a}chtler F. \and Santos L.}
\REVIEW{Phys. Rev. A}{93}{2016}{061603(R)}.

\bibitem{chredh} 
\Name{Kawaguchi Y., Saito H. \and Ueda M.}
\REVIEW{Phys. Rev. Lett.}{96}{2006}{080405}.

\bibitem{rubid1} 
\Name{Gawryluk K., Brewczyk M., Bongs K. \and Gajda M.}
\REVIEW{Phys. Rev. Lett.}{99}{2007}{130401}.

\bibitem{rubid2} 
\Name{\'Swis\l{}ocki T., Gajda M. \and Brewczyk M.}
\REVIEW{Phys. Rev. A}{90}{2014}{063635}.

\bibitem{rubid3} 
\Name{Gawryluk K., Bongs K. \and Brewczyk M.}
\REVIEW{Phys. Rev. Lett.}{106}{2011}{140403}.

\bibitem{opty1} 
\Name{Greiner M., Mandel O., Esslinger T., H{\"a}nsch T.W. \and Bloch I.}
\REVIEW{Nature}{415}{2002}{39}.

\bibitem{opty2} 
\Name{Serwane F., Zurn G., Lompe T., Ottenstein T. B., Wenz A. N. \and Jochim S.}
\REVIEW{Science}{332}{2011}{336}.

\bibitem{2atomy1} 
\Name{Sun B. \and You L.}
\REVIEW{Phys. Rev. Lett.}{99}{2007}{150402}.

\bibitem{2atomy2} 
\Name{Pietraszewicz J., Sowi\'nski T., Brewczyk M., Lewenstein M. \and Gajda M.}
\REVIEW{Phys. Rev. A}{88}{2013}{013608}.

\bibitem{my} 
\Name{O\l{}dziejewski R., G\'orecki W. \and Rz\k{a}\.zewski K.}
\REVIEW{Europhysics Letters}{114}{2016}{46003}.

\bibitem{kontakt} 
\Name{Petrov A., Tiesinga E. \and Kotochigova S.}
\REVIEW{Phys. Rev. Lett.}{109}{2012}{103002}.

\bibitem{rez} 
\Name{Maier T., Ferrier-Barbut I., Kadau H., Schmitt M., Wenzel M., Wink C., Pfau T., Jachymski K. \and Julienne P. S.}
\REVIEW{Phys. Rev. A}{92}{2015}{060702(R)}.

\bibitem{adiabatic} 
\Name{Born M. and Fock V.}
\REVIEW{Zeitschrift f\"ur Physik}{51}{1928}{165}.

\bibitem{landau} 
\Name{Landau L.}
\REVIEW{Phys. Z. Sowjetunion}{2}{1932}{46}.

\bibitem{zener} 
\Name{Zener C.}
\REVIEW{Proc. R. Soc. A.}{137}{1932}{696}.




\end{thebibliography}
\end{document}